\documentclass{revtex4}
\usepackage{graphicx}
\usepackage{fancyhdr}
\usepackage{amsmath}
\def\ignore#1{{}}

\newcommand{\beeq}{\begin{equation}}
\newcommand{\eneq}{\end{equation}}
\newcommand{\beqn}{\begin{eqnarray}}
\newcommand{\eeqn}{\end{eqnarray}}

\def\mybig{\displaystyle \strut }

\def\la{\raise.16ex\hbox{$\langle$}\lower.16ex\hbox{}  }
\def\ra{\raise.16ex\hbox{$\rangle$}\lower.16ex\hbox{} }
\def\go{\rightarrow}

\def\onehalf{ \hbox{$\frac{1}{2}$} }

\def\eff{{\rm eff}}

\def\cF{{\cal F}}
\def\cL{{\cal L}}

\def\SM{{\rm SM}}

\def\KK{{\rm KK}}

\def\psibar{ \psi \kern-.65em\raise.6em\hbox{$-$} }
\def\psibarl{ \psi \kern-.65em\raise.6em\hbox{$-$} \lower.6em\hbox{} }

\def\myfrac#1#2{{\mybig #1\over \mybig #2}}

\usepackage{color}

\begin{document}

~
\vskip -10pt
\title{\Large 126 GeV Higgs boson and universality relations\\
in the $SO(5) \times U(1)$  gauge-Higgs unification}

%

\author{Yutaka Hosotani}
\affiliation{Department of Physics, Osaka University, Toyonaka, Osaka 560-0043, Japan}

\begin{abstract}
The Higgs boson mass $m_H=126\,$GeV in the  $SO(5) \times U(1)$  gauge-Higgs
unification in the Randall-Sundrum space leads to important consequences.
An  universal relation  is found between the Kaluza-Klein (KK) mass scale
$m_{KK}$ and the Aharonov-Bohm phase $\theta_H$ in the fifth dimension;
$m_{KK} \sim 1350\,{\rm GeV}/(\sin \theta_H)^{0.787}$.
The cubic and quartic self-couplings of the Higgs boson become smaller than
those
in the SM, having universal dependence on $\theta_H$.
The decay rates $H \go \gamma \gamma, gg$ are evaluated by summing
contributions
from KK towers.   Corrections coming from KK excited states turn out very
small.
With $\theta_H= 0.1 \sim 0.35$, the mass of the first KK $Z$ is predicted
to be  $2.5 \sim 6 \, $TeV.
\end{abstract}

\maketitle

\thispagestyle{fancy}


\section{Introduction}
The discovery of a Higgs-like boson with $m_H =126 \,$GeV at LHC may give 
a hint for extra dimensions.  
We show \cite{Hosotani2013} that the observed Higgs boson mass  in the 
gauge-Higgs unification scenario 
leads to universal relations among the AB phase $\theta_H$, the KK mass $m_\KK$,
the Higgs self couplings, and the KK $Z$ boson mass $m_{Z^{(1)}}$,  independent
of the details of the model.

The gauge-Higgs unification scenario is  predictive.
As a result of the Hosotani mechanism \cite{YH1, Davies1, KLY, HHHK, Hatanaka1998} 
the Higgs boson mass emerges at the quantum level without  being afflicted 
with divergence.   The Higgs couplings to
the KK towers of quarks and $W/Z$ bosons have a distinctive feature that
their signs alternate in the KK level, significant departure from other
extra dimensional models such as UED models.  As a consequence contributions of
KK modes to the decay rate $\Gamma ( H \go \gamma \gamma)$ turn
out very small.  Surprisingly the gauge-Higgs unification gives nearly the 
same phenomenology at low energies as the standard model (SM).

The gauge-Higgs unification can be confirmed by finding the  KK $Z$ boson 
in the range $2.5 \sim 6 \, $TeV and by determining the Higgs self couplings
and Yukawa couplings at LHC and ILC.

\section{$SO(5) \times U(1)$  gauge-Higgs unification in RS}

\def\mybox#1{\fbox{\rule{0cm}{8pt}\rule{2mm}{0pt}#1\rule{2mm}{0pt}}}
\def\mathbox#1{\hbox{\fbox{\rule{0cm}{8pt}\rule{1mm}{0pt}#1\rule{1mm}{0pt}}}}

The model is given by $SO(5) \times U(1)$ gauge theory in the  Randall-Sundrum (RS) 
warped space
\beeq
ds^2 =  e^{-2\sigma(y)}\eta_{\mu\nu} dx^\mu dx^\nu + dy^2 
\label{metric1}
\eneq
where $\eta_{\mu\nu} =\textrm{diag}(-1,1,1,1)$,
$\sigma(y)=\sigma(y+2L) = \sigma(-y)$, and $\sigma(y)=k|y|$ for $|y|\leq L$.
The RS space is viewed as bulk AdS space ($0 < y < L$) with AdS curvature $-6k^2$ 
sandwiched by the Planck brane at $y=0$ and the TeV brane at $y=L$.  
The $SO(5) \times U(1)$ model was proposed by Agashe et al \cite{ACP, MSW}.
It has been elaborated in refs.~\cite{HOOS, HNU}, and a concrete realistic model 
has been formulated in ref.~\cite{Hosotani2013}.
The schematic view of  the gauge-Higgs unification is given below.

\beqn
&&\hskip -1.cm
\mathbox{5D $A_M$} ~
\begin{cases}
~\mathbox{four-dim. components $A_\mu$} 
&\in \mathbox{4D gauge fields  $\gamma, W, Z$}\cr
\noalign{\kern 25pt}
~\mathbox{extra-dim. component $A_y$}
&\in \mathbox{4D Higgs field $H$}
\end{cases}  \cr
\noalign{\kern 10pt}
&&\hskip 7.0cm
\sim \mathbox{AB phase $\theta_H$} \hbox{ ~in extra dim.}\cr
\noalign{\kern 15pt}
&&\hskip 4.5cm
\hbox{\bf  Hosotani mechanism}  \hskip .3cm \Downarrow  \cr
\noalign{\kern 15pt}
&&\hskip 5.5cm
\underline{\hbox{\bf Dynamical EW symmetry breaking}\strut}
\nonumber
\eeqn

The 5D Lagrangian density consists of 
\begin{align}
\cL = & ~ \cL_{\rm bulk}^{\rm gauge} (A, B) 
+ \cL_{\rm bulk}^{\rm fermion} (\Psi_a, \Psi_F, A, B)  \cr
\noalign{\kern 5pt}
&+\cL_{\rm brane}^{\rm fermion} (\hat \chi_\alpha, A, B)
+ \cL_{\rm brane}^{\rm scalar} (\hat \Phi, A, B)
+ \cL_{\rm brane}^{\rm int} (\Psi_a,\hat \chi_\alpha,\hat \Phi)  ~.
\label{Lagrangian1}
\end{align}
$SO(5)$ and $U(1)_X$ gauge fields are denoted by $A_M$ and $B_M$, respectively.
The two associated gauge coupling constants are  $g_A$ and $g_B$.
Two quark multiplets and two lepton multiplets $\Psi_a$ are introduced in the vector 
representation of $SO(5)$ in each generation,  whereas $n_F$ extra fermion multiplets 
$\Psi_F$  are introduced in the spinor representation.
These bulk fields obey the orbifold boundary conditions at $y_0=0$ and $y_1=L$ given by
\begin{align}
&\begin{pmatrix}    A_\mu \cr A_y    \end{pmatrix}  (x,y_j-y)
= P_j  \begin{pmatrix}    A_\mu \cr - A_y  \end{pmatrix}  (x,y_j+y) P_j^{-1} ,  \cr
\noalign{\kern 5pt}
&\begin{pmatrix}    B_\mu \cr B_y    \end{pmatrix}  (x,y_j-y)
= \begin{pmatrix}    B_\mu \cr - B_y  \end{pmatrix}  (x,y_j+y) ~ , \cr
\noalign{\kern 5pt}
&~~ \Psi_a (x,y_j-y) = P_j \Gamma^5 \Psi_a (x,y_j+y) ~ , ~~
 \Psi_F (x,y_j-y) = (-1)^j P_j^{\rm sp} \Gamma^5 \Psi_F (x,y_j+y) ~ , \cr
\noalign{\kern 5pt}
& ~~ P_j =\textrm{diag} \, (-1,-1,-1,-1,1)~, ~~~ 
P_j^{\rm sp} =\textrm{diag} \, (1,1,-1,-1) ~.
\label{BC1}
\end{align}
The orbifold boundary conditions break $SO(5) \times U(1)_X$ to 
$SO(4) \times U(1)_X \simeq SU(2)_L \times SU(2)_R \times U(1)_X$.

The brane  interactions are invariant under $SO(4) \times U(1)_X$.
The brane scalar $\hat \Phi$ is in the $({\bf 1, 2})_{-1/2}$ representation of 
$[SU(2)_L , SU(2)_R]_{U(1)_X}$.  It spontaneously breaks $SU(2)_R \times U(1)_X$
to $U(1)_Y$ by non-vanishing $\la \hat \Phi \ra$ whose magnitude is supposed to be
much larger than the KK scale $m_\KK$.  
At this stage the residual gauge symmetry is $SU(2)_L \times U(1)_Y$.  
Brane fermions $\hat \chi_\alpha$ are introduced in the $({\bf 2, 1})$ representation.
The quark-lepton vector multiplets $\Psi_a$ are decomposed into $({\bf 2, 2}) + ({\bf 1, 1})$.
The $({\bf 2, 2})$ part of $\Psi_a$, 
$\hat \chi_\alpha$ in $({\bf 2,1})$  and $\hat \Phi$ in $({\bf 1,2})$ form 
$SO(4) \times U(1)_X$ invariant brane interactions.  With $\la \hat \Phi \ra \not= 0$
they yield mass terms.  The resultant spectrum of massless fermions is the same as 
in the SM.  All exotic fermions become heavy, acquiring masses of $O(m_\KK)$.  
Further with brane fermions all anomalies associated with gauge fields of $SO(4) \times U(1)_X$ 
are cancelled.\cite{HNU}

With the orbifold boundary conditions (\ref{BC1}) there appear four zero modes of $A_y$
in the components $(A_y)_{a5} = - (A_y)_{5a}$ ($a=1, \cdots, 4$).  They form an $SO(4)$ vector,
or an $SU(2)_L$ doublet,  corresponding to the Higgs doublet in the SM.
The AB phase is defined with these zero modes by
\beeq
e^{i \Theta_H/2} \sim P \exp \bigg\{ i g_A \int_0^L dy \, A_y \bigg\} ~.
\label{ABphase1}
\eneq
At the tree level the value of the AB phase $\Theta_H$ is not determined, as it gives vanishing
field strengths.  At the quantum level its effective potential $V_\eff$ becomes non-trivial.  
The value of $\Theta_H$ is  determined by the location of the minimum of $V_\eff$.  
This is the Hosotani mechanism and induces dynamical gauge symmetry 
breaking.    It leads to gauge-Higgs unification, resolving the gauge-hierarchy 
problem.\cite{Hatanaka1998}
Without loss of generality  one can assume that $(A_y)_{45}$ component develops 
a non-vanishing expectation value.
Let us denote the corresponding component of $\Theta_H$  by $\theta_H$.
If $\theta_H$ takes a non-vanishing value, the electroweak symmetry breaking takes place.

\section{$V_\eff(\theta_H)$ and $m_H$}

Given the matter content one can evaluate $V_\eff (\theta_H)$ at the one loop level
unambiguously.  The $\theta_H$ dependent part of  $V_\eff (\theta_H)$ is finite,
being free from divergence. 
$V_\eff ( \theta_H)$ depends on several parameters of the theory; 
$V_\eff = V_\eff (\theta_H; \xi, c_t, c_F, n_F, k, z_L)$ where $\xi$ is the gauge 
parameter in the generalized $R_\xi$ gauge, $c_t$  and $c_F$ are the  bulk mass
parameters of the top  and extra fermion multiplets,   $n_F$ is the number of the 
extra fermion multiplets, and $k, z_L$ are parameters
specifying the RS metric  (\ref{metric1}). 
Given these parameters, $V_\eff$ is fixed, and the location of the global minimum
of $V_\eff (\theta_H)$, $\theta_H^{\rm min}$, is determined.

With $\theta_H^{\rm min}$ determined, $m_Z$, $g_w$, $\sin^2 \theta_W$
are determined from  $g_A, g_B, k, z_L$ and $\theta_H^{\rm min}$.
The top mass $m_t$ is determined from $c_t, k, z_L, \theta_H^{\rm min}$, whereas
the Higgs boson mass  $m_H$ is given by
\beeq
m_H^2 = \frac{1}{f_H^2} \frac{d^2 V_\eff}{d \theta_H^2} \bigg|_{\rm min},
~~~~
f_H = \frac{2}{g_w} \sqrt{ \frac{k}{L(z_L^2 -1)} } ~.
\label{Higgs1}
\eneq
Let us take $\xi=1$. Then the theory has seven  parameters
$\{ g_A, g_B, k, z_L, c_t, c_F, n_F \}$.  Adjusting theses parameters,
we reproduce the values of  five observed quantities $\{ m_Z, g_w, \sin^2 \theta_W, m_t, m_H\}$.
This leaves two parameters, say $z_L$ and $n_F$,  free.
Put  differently, the value of $\theta_H^{\rm min}$  is determined as a function of
$z_L$ and $n_F$;  $\theta_H^{\rm min} = \theta_H (z_L, n_F)$.    
We comment that contributions from other light quark/lepton multiplets to $V_\eff$
are negligible.  

$V_\eff (\theta_H)$ in the absence of the extra fermions ($n_F=0$) was evaluated
in refs.~\cite{HOOS, HTU1}.  It was found there that the global minima naturally appear 
at $\theta_H = \pm \onehalf \pi$ at which the Higgs boson becomes absolutely stable.
It is due to the emergence of the $H$ parity invariance.\cite{HKT, HTU1}
In particular the Higgs trilinear couplings to $W$, $Z$, quarks and leptons are all
proportional to $\cos \theta_H$ and  vanish at 
$\theta_H = \pm \onehalf \pi$.\cite{SH1, HS2, Sakamura1, Giudice2007, HK, Hasegawa}

This, however, conflicts with the observation of an unstable Higgs boson at LHC.
To have an unstable Higgs boson the $H$ parity invariance must be broken,
which is most easily achieved by introducing extra fermion multiplets $\Psi_F$
in the spinor representation of $SO(5)$ in the bulk.\cite{Hosotani2013} 

\begin{figure}[h]
\centering
\includegraphics[height=45mm]{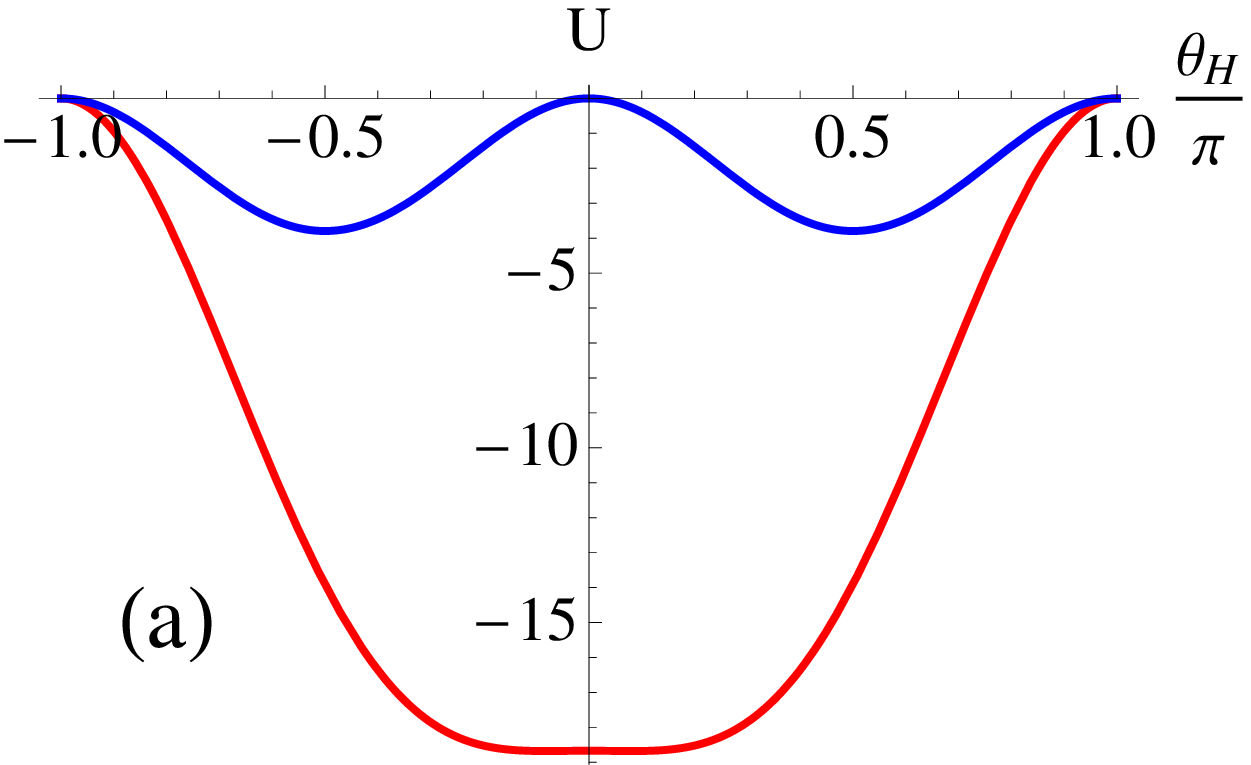}
\qquad
\includegraphics[height=45mm]{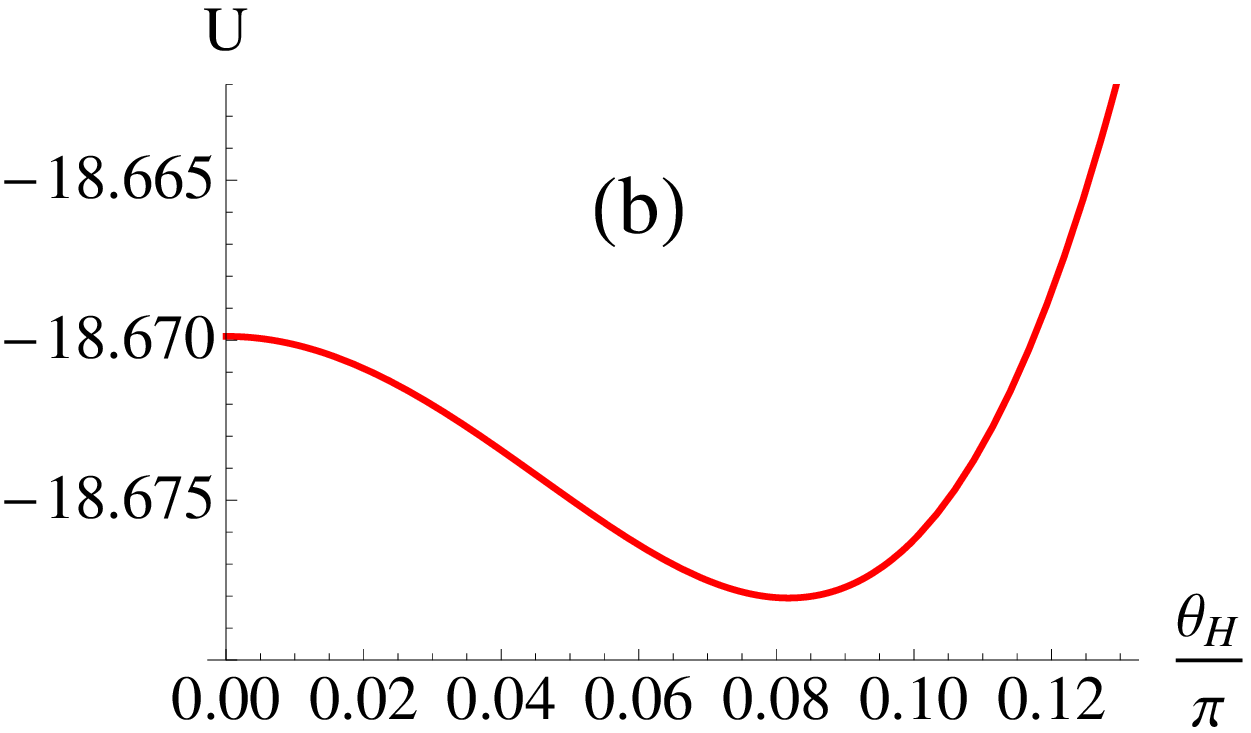}\\
\caption{The effective potential $V_\eff (\theta_H)$ for $z_L=10^7$.  
$U =  16 \pi^6 m_\KK^{-4} V_\eff$ is plotted.
The red curves are for $n_F=3$ with $m_H = 126\,$GeV. $V_\eff$ has minima at
$\theta_H = \pm 0.258$ and $m_\KK = 3.95\,$TeV.  
The blue curve is for $n_F=0$ in which  case $m_H = 87.9\,$GeV and $m_\KK = 993\,$GeV. 
}
\label{effV_figure}
\end{figure}

Let us take $n_F=3, z_L=e^{kL} = 10^7$ as an example.  $\{ g_w, \sin^2 \theta_W \}$ are 
related to $\{ g_A, g_B \}$ by
\beeq
g_w = \frac{g_A}{\sqrt{L}} ~~,~~ \tan \theta_W = \frac{g_B}{\sqrt{g_A^2 + g_B^2}} ~~,
\label{gw1}
\eneq
where $z_L= e^{kL}$. 
The observed values of  $\{ m_Z, g_w, \sin^2 \theta_W, m_t, m_H\}$ are reproduced with 
$k= 1.26 \times 10^{10} \,$GeV, $c_t = 0.330$, $c_F = 0.353$ for which the minima of $V_\eff$ 
are found at $\theta_H = \pm 0.258$.  The KK mass scale is $m_\KK = \pi k z_L^{-1}= 3.95\,$TeV. 
$V_\eff (\theta_H)$ is depicted in Fig.~{\ref{effV_figure}}  with red curves.  
For comparison $V_\eff$ in the case of $n_F=0$ is also plotted with a blue curve.
When $n_F=0$ and $z_L=10^7$, the minima are located at $\theta_H = \pm \onehalf \pi$.
The observed values of  $\{ m_Z, g_w, \sin^2 \theta_W, m_t\}$ are reproduced with 
$k=3.16 \times 10^{9} \,$GeV and $c_t = 0.345$.  In this case the Higgs boson mass 
determined by (\ref{Higgs1})  becomes $m_H = 87.9\,$GeV, and  $m_\KK = 993\,$GeV. 
One can see how the position of the minima is shifted from  $\theta_H = \pm \onehalf \pi$
to $\theta_H = \pm 0.082 \pi = \pm 0.258$ by the introduction of the extra fermions.

\section{Universality}
\begin{figure}[htb]
\centering
\includegraphics[height=50mm]{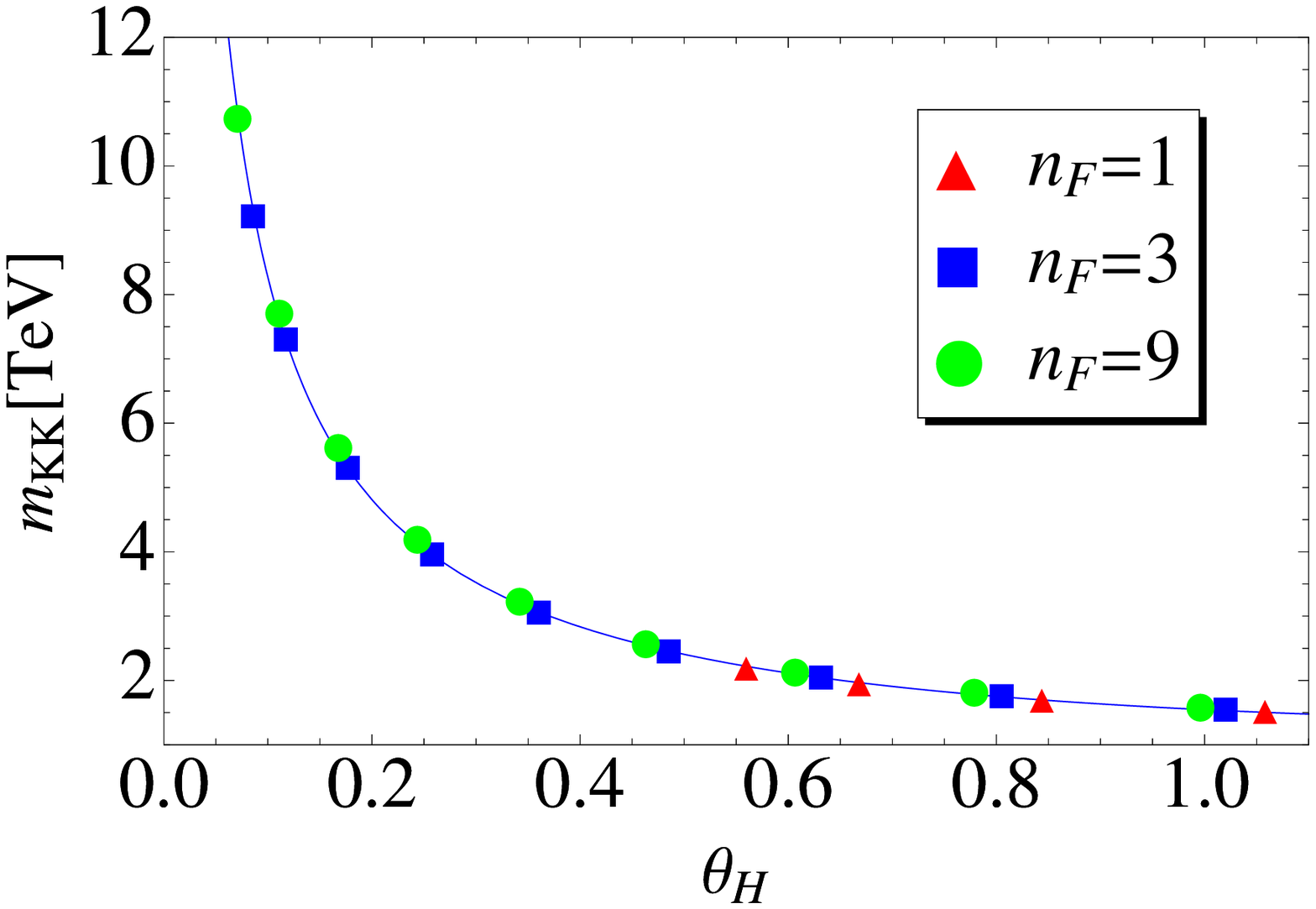}\\
\includegraphics[height=45mm]{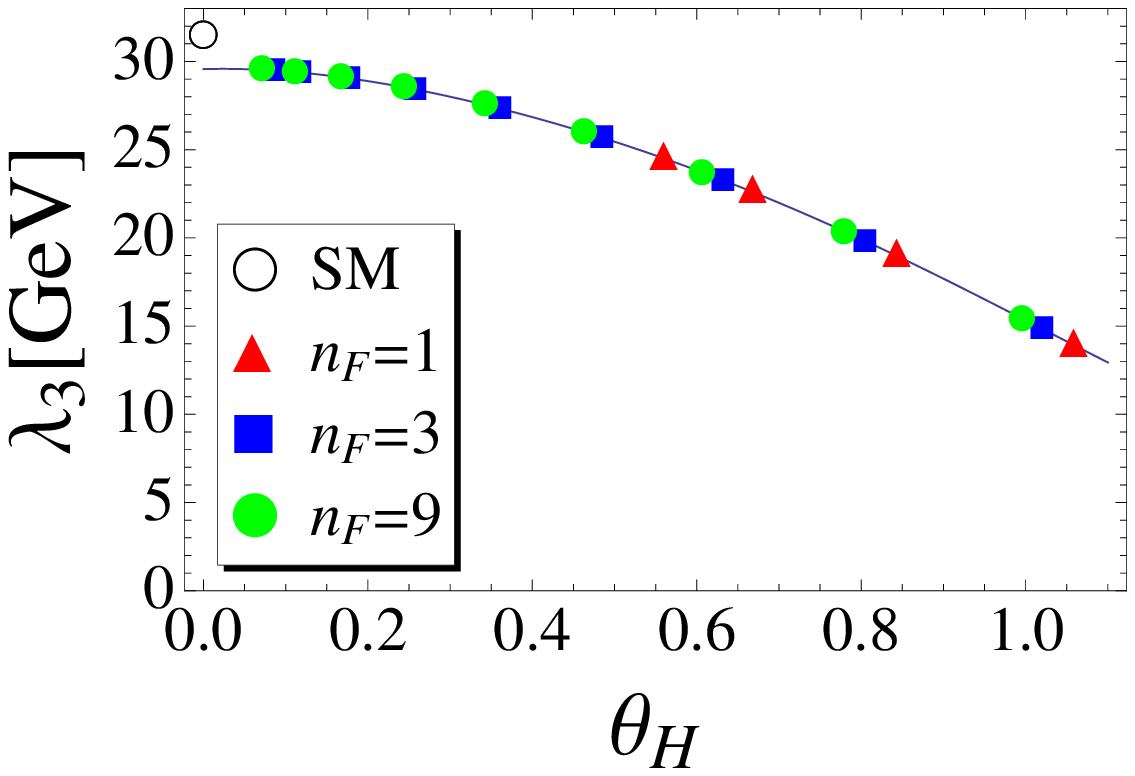}\quad
\includegraphics[height=45mm]{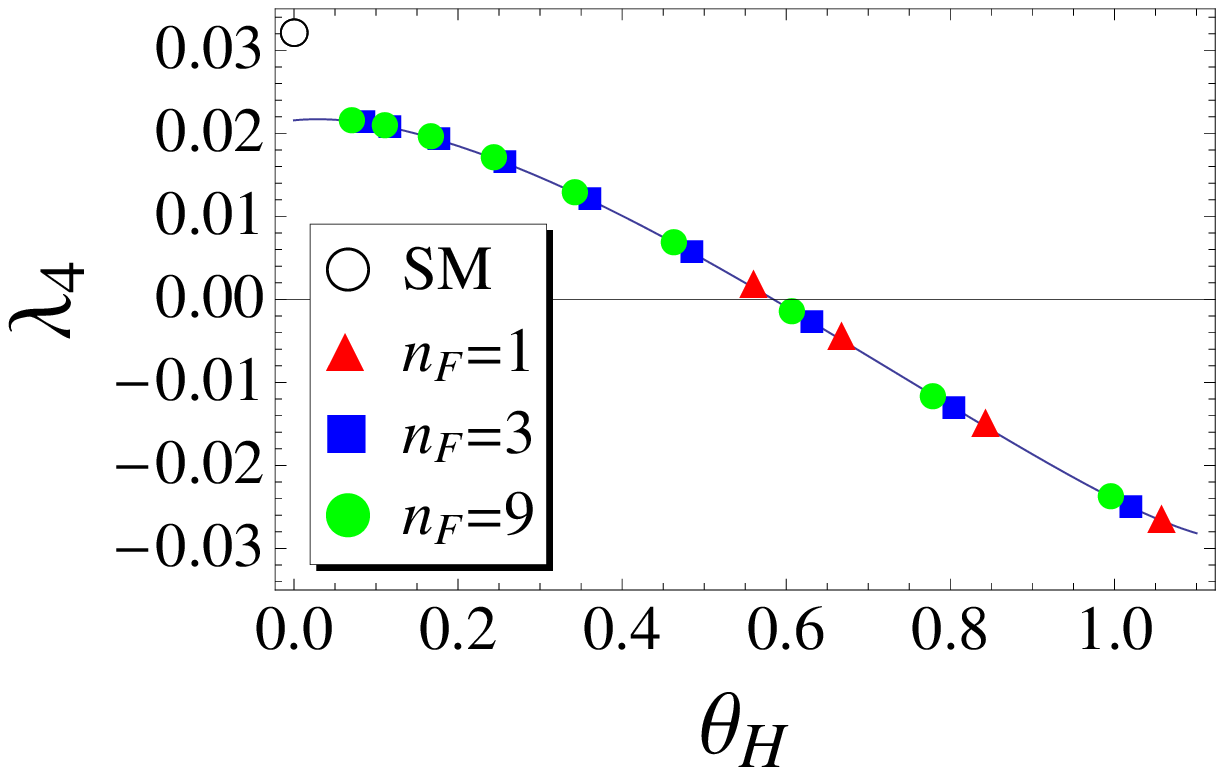}
\caption{Universality relations. [Top] KK scale $m_\KK (\theta_H)$.
[Bottom left] Higgs cubic self-coupling $\lambda_3(\theta_H)$.
[Bottom right] Higgs quartic self-coupling $\lambda_4(\theta_H)$.
The universality, independent of $n_F$, is seen in all relations.
} 
\label{universal}
\end{figure}

As explained above, the AB phase $\theta_H ( = \theta_H^{\rm min})$  is determined
as a function of $z_L$ and $n_F$; $\theta_H (z_L, n_F)$.   
The KK mass scale $m_\KK = \pi k z_L^{-1}$
is also determined as a function of $z_L$ and $n_F$; $m_\KK (z_L, n_F)$.
The relation between them is plotted  for $n_F = 1,3,9$ in the top figure in Fig.~\ref{universal}.
One sees that all points fall on one universal curve to good accuracy, independent of $n_F$.

Similarly one can evaluate the cubic ($\lambda_3$) and quartic  ($\lambda_4$) 
self-couplings of the Higgs boson $H$ by expanding $V_\eff [\theta_H + (H/f_H)]$ 
around the minimum  in a power series in $H$.  They are depicted in the bottom figure in Fig.~\ref{universal}.
Although the shape of $V_\eff (\theta_H)$ heavily depends on $n_F$,
the relations $\lambda_3(\theta_H)$ and $\lambda_4(\theta_H)$ turn out universal, 
independent of $n_F$.

It is rather surprising that there hold universal relations among $\theta_H$, $m_\KK$,
$\lambda_3$ and $\lambda_4$.
Once $\theta_H$ is determined from one source of observation, then many other physical quantities
are fixed and predicted.  
The gauge-Higgs unification gives many definitive predictions to be tested by experiments.
We tabulate values of various quantities determined from $m_H=126\,$GeV with given 
$z_L$ for $n_F=3$ in Table \ref{table1}.  The relation between $\theta_H$ and $m_\KK$
is well summarized with
\beeq
m_\KK \sim \myfrac{1350 \, {\rm GeV}}{( \sin \theta_H )^{0.787}} ~.
\label{mKKtheta1}
\eneq

\begin{table}[htb]
\caption{Values of the various quantities with given $z_L$ for $n_F=3$.
$m_{Z^{(1)}}$ and $m_{F^{(1)}}$ are masses of the first KK $Z$ boson and the
lowest mode of the extra fermion multiplets.
Relations among $\theta_H$, $m_\KK$ and $m_{Z^{(1)}}$ are universal, 
independent of $n_F$.}
\begin{center}
\renewcommand{\arraystretch}{1.3}
\begin{tabular}{|c||c|c|c||c|}
\hline
$z_L$  & $\theta_H$ & ~$m_{\mbox {{\tiny KK}}}$~ 
    &~$m_{Z^{(1)}}$~  &~$ m_{F^{(1)}} $~ \\
\hline
~~$10^8$~~   & ~~$0.360$~~ & $ ~~3.05${\small ~TeV}~~  &~~$ 2.41 ${\small ~TeV}~~ 
&~~$0.668${\small ~TeV}~~ \\
$10^7$  & $0.258$ & $ 3.95$  &$3.15  $  &$0.993$\\
$10^6$  & $ 0.177$ & $5.30$  &$ 4.25 $ &$1.54~$\\
$10^5$ & $0.117$ & $ 7.29$   &$ 5.91 $ &$2.53~$\\
\hline
\end{tabular}
\end{center}
\label{table1}
\end{table}

\section{$H \go \gamma \gamma, gg$}

In the gauge-Higgs unification all of the 3-point couplings of $W$, $Z$, quarks and
leptons to the Higgs boson $H$  at the tree level are suppressed by a common factor 
$\cos \theta_H$ compared with those 
in the SM.\cite{SH1, HS2, Sakamura1, Giudice2007, HK, Hasegawa}
The decay of the Higgs boson to two photons  goes through loop diagrams in which
$W$ boson, quarks, leptons, extra fermions and their KK excited states run.  

The decay rate  $\Gamma [ H \go \gamma \gamma ]$ is given by
\beqn
&&\hskip -1.cm
\Gamma ( H \go \gamma \gamma ) = \frac{\alpha^2 g_w^2}{1024 \pi^3}
\frac{m_H^3}{m_W^2} ~ \big| \cF_{\rm total} \big|^2 ~,  \cr
\noalign{\kern 5pt}
&&\hskip -1.cm
\cF_{\rm total} =  \cF_W + \frac{4}{3} \cF_{\rm top}
+ \left(2 (Q_X^{(F)})^2 + \onehalf\right)  n_F \cF_{F} ~, \cr
\noalign{\kern 5pt}
&&\hskip -1.cm
\cF_W = \cos \theta_H \sum_{n=0}^\infty I_{W^{(n)}}
\frac{m_W}{m_{W^{(n)}}} \,  F_1 ( \tau_{W^{(n)}} ) ~,~~
I_{W^{(n)}} = \frac{g_{HW^{(n)} W^{(n)}} }{ g_w m_{W^{(n)}} \cos \theta_H }~,   \cr
\noalign{\kern 5pt}
&&\hskip -1.cm
\cF_{\rm top} =  \cos \theta_H \sum_{n=0}^\infty  I_{t^{(n)}}
 \frac{m_t}{m_{t^{(n)}}} \,  F_{1/2} ( \tau_{t^{(n)}} ) ~,~~
 I_{t^{(n)}} = 
\frac{y_{t^{(n)}} }{ y_t^\SM \cos \theta_H } ~,    \cr
\noalign{\kern 5pt}
&&\hskip -1.cm
\cF_F = \sin \onehalf \theta_H \sum_{n=1}^\infty  I_{F^{(n)}} 
 \frac{m_t}{m_{F^{(n)}}} \,  F_{1/2} ( \tau_{F^{(n)}} ) ~,~~
 I_{F^{(n)}} = \frac{y_{F^{(n)}}}{y_t^\SM \sin \onehalf \theta_H} ~,   
\label{Hdecay1}
\eeqn
where $W^{(0)} = W$, $t^{(0)} =t$, $\tau_{a} = 4 m_{a}^2/m_H^2$.
The functions $F_1(\tau)$ and $F_{1/2}(\tau)$ are defined in Ref.~\cite{HiggsHunter},
and $F_1(\tau) \sim 7$ and $F_{1/2} (\tau) \sim - \frac{4}{3}$ for $\tau \gg 1$. 
$Q_X^{(F)}$ is the $U(1)_X$ charge of the extra fermions.
$I_{W^{(0)}}$ and $I_{t^{(0)}}$ are $ \sim 1$.

\begin{figure}[htb]
\centering
\includegraphics[height=55mm]{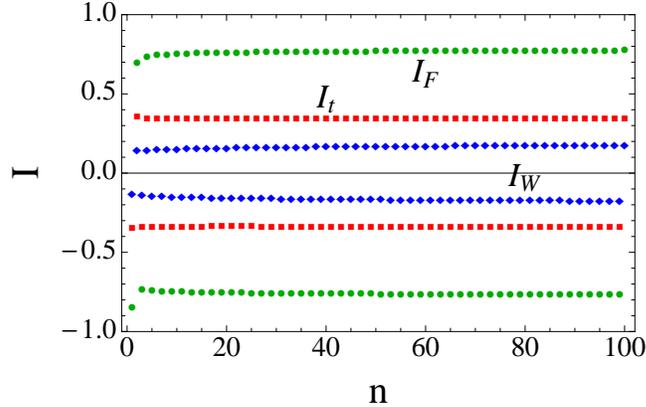}
\caption{$I_{W^{(n)}}$, $I_{t^{(n)}}$, and $I_{F^{(n)}}$ for 
$n_F=3$, $Q_X^{(F)} = 0$ and   $\theta_H= 0.360$ ($z_L=10^8$) in the range $1 \le n \le 100$.
$I_{W^{(0)}} = 1.004$ and $I_{t^{(0)}} = 1.012$.
} 
\label{loop}
\end{figure}

In Fig.~\ref{loop}, $I_{W^{(n)}}$, $I_{t^{(n)}}$, and $I_{F^{(n)}}$ are plotted.
One sees that the values  of these $I$'s alternate in sign as $n$ increases, which 
gives sharp contrast to the UED models.  
\beeq
I_{W^{(n)}} \sim (-1)^n I_{W}^\infty ~~,~~
I_{t^{(n)}} \sim (-1)^n I_{t}^\infty ~~,~~
I_{F^{(n)}} \sim (-1)^n I_{F}^\infty  \qquad {\rm for~} n \gg 1 
\label{Hdecay2}
\eneq
up to $(\ln n)^p$ corrections.
This is special to the gauge-Higgs unification
models.  It has been known in the models in flat space as well.\cite{MaruOkada, Falkowski2008}
As a consequence of the destructive interference due to the alternating sign, 
the infinite sums in the rate (\ref{Hdecay1}) converges rapidly.  There appears no divergence.

Let $\cF_{W \, \rm only}$ and $\cF_{t \, {\rm only}}$ be the contributions of 
$W=W^{(0)}$ and $t=t^{(0)}$ to $\cF_{\rm total}$. 
The numerical values of the amplitudes $\cF$'s are tabulated in Table \ref{table2} for $n_F=3$.
It is seen that contributions of KK states to the amplitude are small.  
The dominant effect  for the decay amplitude is the suppression factor $\cos \theta_H$.

\begin{table}[tbh]
\caption{Values of the amplitudes $\cF$'s in (\ref{Hdecay1}) for $n_F=3$ and $Q_X^{(F)} = 0$.
}
\begin{center}
\renewcommand{\arraystretch}{1.4}
\begin{tabular}{|cc|rr|rr|}
\hline
\multicolumn{2}{|c}{$\theta_H$}    
& \multicolumn{2}{|c}{~0.117~} & \multicolumn{2}{|c|}{~0.360~} \\
\multicolumn{2}{|c}{$z_L$}    
& \multicolumn{2}{|c}{$10^5$} & \multicolumn{2}{|c|}{$10^8$} \\
\hline
\quad $\cF_{W \, \rm only}$ &&8.330 &&7.873 & \\
& $\cF_{W} /\cF_{W \, {\rm only}}$ &&0.9996 \quad &&0.998 \quad \\
\hline
\quad $\cF_{t \, \rm only}$ &&~~-1.372 &&-1.305 & \\
& $\cF_{t} /\cF_{t \, {\rm only}}$ &&0.998 \quad &&0.990 \quad \\
\hline
& $\cF_{F} /\cF_{t \, {\rm only}}$ &&~~-0.0034 \quad &&~~-0.033 \quad \\
\hline
\multicolumn{2}{|c|}{$\cF_{\rm total}$}   &6.508 &&6.199 & \\
\multicolumn{2}{|c|}{\quad $\cF_{\rm total}/(\cF_{W \, {\rm only}} + \frac{4}{3} \cF_{t \, {\rm only}})$ \quad} 
      &&1.001 \quad &&1.011 \quad \\
\hline
\end{tabular}
\end{center}
\label{table2}
\end{table}

All Higgs couplings $HWW, HZZ, Hc \bar c, Hb \bar b, H \tau \bar \tau$ are suppressed by a factor
$\cos \theta_H$ at the tree level.   The corrections to $\Gamma [H \go \gamma \gamma]$ and 
$\Gamma [H \go g g]$ due to KK states amount only to 0.2\% (2\%) for $\theta_H = 0.117 (0.360)$.
Hence we conclude
\beqn
\hbox{branching fraction:} && B ( H \go j) \sim B^{\SM} ( H \go j) \cr
\noalign{\kern 5pt}
&&\hskip 1.cm   j = WW, ZZ, \gamma \gamma, gg , b\bar b, c \bar c, \tau \bar\tau, \cdots \cr
\noalign{\kern 10pt}
\gamma \gamma ~ \hbox{production rate:} &&  \sigma^{\rm prod} (H) \cdot B(H \go \gamma \gamma) 
\sim ({\rm SM}) \times \cos^2 \theta_H ~.
\label{rate1}
\eeqn
The signal strength in the $\gamma \gamma$ production relative to the SM is about $\cos^2 \theta_H$.
It is about 0.99 (0.91) for $\theta_H = 0.1 ~  (0.3)$.  
This contrasts to the prediction in the UED models in which the contributions of KK states 
can add up in the same sign to sizable amount.\cite{ued}

\vskip 1cm

\section{Signals of gauge-Higgs unification}
There are several constraints to be imposed on the gauge-Higgs unification.

(i) For the consistency with the $S$ parameter, we need $\sin \theta_H < 0.3$.\cite{ACP}

(ii) The tree-level unitarity requires $\theta_H < 0.5$.\cite{unitarity}

(iii) $Z'$ search at Tevatron and LHC.  The first KK $Z$ corresponds to $Z'$.
No signal has been found so far, which implies that $m_{Z^{(1)}} > 2 \,$TeV.
With the universality relations in Sec. IV it requires $\theta_H < 0.4$.

(iv) In ref.~\cite{HTU2} the consistency with other precision measurements
such as the $Z$ boson decay and the forward-backward asymmetry on the $Z$ resonance
has been investigated when $n_F=0$.  Reasonable agreement was found for $m_\KK > 1.5\,$TeV.
We need to reanalyze in the case $n_F \ge 1$.


All of those constraints above point $\theta_H < 0.4$.  When $\theta_H$ is very small,
the KK mass scale $m_\KK$ becomes very large and it becomes very difficult
to distinguish the gauge-Higgs unification from the SM.  
The  range of interest  is $0.1 < \theta_H < 0.35$, which can be explored at LHC  with
an increased energy 13 or 14 TeV.  The gauge-Higgs unification predicts the following
signals.

(1) The first KK $Z$ should be found at $m_\KK = 2.5 \sim 6 \,$TeV for 
$\theta_H = 0.35 \sim 0.1$.

(2) The Higgs self-couplings should be smaller than those in the SM.
$\lambda_3$ ($\lambda_4$) should be $10 \sim 20$\% ($30 \sim 60$\%)
smaller for $\theta_H = 0.1 \sim 0.35$, according to the universality relations.
This should be explored at ILC.

(3)  The lowest mode ($F^{(1)}$) of the KK tower of the extra fermion $\Psi_F$ should be discovered
at LHC.  Its mass depends on both $\theta_H$ and $n_F$. For $n_F=3$, the mass is predicted to be
$m_{F^{(1)}} = 0.7 \sim 2.5 \,$TeV for $\theta_H = 0.35 \sim 0.1$.

\section{For the Future}

The $SO(5) \times U(1)$ gauge-Higgs unification model of ref.~\cite{Hosotani2013} 
has been  successful so far.  Yet further elaboration may be necessary.

(1) Flavor mixing has to be incorporated to explore flavor physics.\cite{flavor}

(2) It is curious to generalize the model to incorporate SUSY.  The Higgs boson mass becomes
smaller than in non-SUSY model.  $m_H = 126\,$GeV should give information about
SUSY breaking scales.\cite{Hatanaka2012}

(3) The orbifold boundary conditions $(P_0, P_1)$ in (\ref{BC1}) have been given by hand so far.
It is desirable to have dynamics which determine the boundary conditions.\cite{YHgut2003, HHK2004}

(4) Not only electroweak interactions but also strong interactions should be
integrated in the form of grand gauge-Higgs unification.\cite{YamashitaGUT}


\begin{acknowledgments}
This work was supported in part 
by  scientific grants from the Ministry of Education and Science, 
Grants No.\ 20244028, No.\ 23104009 and  No.\ 21244036.
\end{acknowledgments}

\bigskip 

\def\jnl#1#2#3#4{{#1}{\bf #2} (#4) #3}

\def\Zphys{{\em Z.\ Phys.} }
\def\jssc{{\em J.\ Solid State Chem.\ }}
\def\jpsJ{{\em J.\ Phys.\ Soc.\ Japan }}
\def\ptps{{\em Prog.\ Theoret.\ Phys.\ Suppl.\ }}
\def\PTP{{\em Prog.\ Theoret.\ Phys.\  }}
\def\JMP{{\em J. Math.\ Phys.} }
\def\NPB{{\em Nucl.\ Phys.} B}
\def\NP{{\em Nucl.\ Phys.} }
\def\PLB{{\it Phys.\ Lett.} B}
\def\PL{{\em Phys.\ Lett.} }
\def\PRL{\em Phys.\ Rev.\ Lett. }
\def\PRB{{\em Phys.\ Rev.} B}
\def\PRD{{\em Phys.\ Rev.} D}
\def\PRe{{\em Phys.\ Rep.} }
\def\AP{{\em Ann.\ Phys.\ (N.Y.)} }
\def\RMP{{\em Rev.\ Mod.\ Phys.} }
\def\ZPC{{\em Z.\ Phys.} C}
\def\SCI{\em Science}
\def\CMP{\em Comm.\ Math.\ Phys. }
\def\MPLA{{\em Mod.\ Phys.\ Lett.} A}
\def\IJMPA{{\em Int.\ J.\ Mod.\ Phys.} A}
\def\IJMPB{{\em Int.\ J.\ Mod.\ Phys.} B}
\def\EPJC{{\em Eur.\ Phys.\ J.} C}
\def\PR{{\em Phys.\ Rev.} }
\def\JHEP{{\em JHEP} }
\def\JCAP{{\em JCAP} }
\def\cmp{{\em Com.\ Math.\ Phys.}}
\def\JPA{{\em J.\  Phys.} A}
\def\JPG{{\em J.\  Phys.} G}
\def\NJP{{\em New.\ J.\  Phys.} }
\def\CQG{\em Class.\ Quant.\ Grav. }
\def\ATMP{{\em Adv.\ Theoret.\ Math.\ Phys.} }
\def\ibid{{\em ibid.} }

\renewenvironment{thebibliography}[1]
         {\begin{list}{[$\,$\arabic{enumi}$\,$]}  
         {\usecounter{enumi}\setlength{\parsep}{0pt}
          \setlength{\itemsep}{0pt}  \renewcommand{\baselinestretch}{1.0}
          \settowidth
         {\labelwidth}{#1 ~ ~}\sloppy}}{\end{list}}

\def\reftitle#1{}                

\end{document}